\documentclass{aa}  

\usepackage{graphicx}
\usepackage{subcaption}
\usepackage{color}
\usepackage{ulem}
\usepackage{txfonts}
\usepackage{xcolor}
\usepackage{natbib}

\usepackage{lipsum}
\usepackage{subcaption}         
\usepackage{lscape}             
\usepackage{placeins}           

\begin{document} 

\newcommand{\psr}{J2022$+$3842}
\newcommand{\gname}{G76.9$+$1.0}
\newcommand{\hii}{H{\sc ii}}
\newcommand{\hi}{H{\sc i}}
\newcommand{\pwn}{the PWN}
\newcommand{\Pwn}{The PWN}
\newcommand{\uv}{\textit{u-v}}
\newcommand{\perbeam}{beam$^{-1}$}
\newcommand{\JyPerBeam}{$\,{\rm Jy\,beam^{-1}}$}
\newcommand{\mJyPerBeam}{$\,{\rm mJy\,beam^{-1}}$}
\newcommand{\uJyPerBeam}{$\,\mu{\rm Jy\,beam^{-1}}$}
\newcommand{\ergPerSecond}{\,erg\ s$^{-1}$}
\newcommand{\ergcmPerSecond}{\,erg\ cm$^{-2}$ s$^{-1}$}
\newcommand{\Msun}{$M_{\odot}$}
\newcommand{\uG}{$\,\mu$G}

\newcommand{\cm}{\,cm}

\newcommand{\hour}{^{\rm h}}
\newcommand{\minute}{^{\rm m}}
\newcommand{\second}{^{\rm s}}
\newcommand{\fsecond}{\fs}
\newcommand{\degrees}{^\circ}
\newcommand{\Chandra}{\texttt{Chandra}}

\newcommand{\task}[1]{\texttt{#1}}
\newcommand{\RMu}{\,rad\,{m$^{-2}$}}

\newcommand{\arcs}{\mbox{$^{\prime\prime}$}}
\newcommand{\arcm}{\mbox{$^{\prime}$}}
   \title{Radio Study of G76.9+1.0 Pulsar Wind Nebula}

   \author{Haotian Qiu
          \inst{1} \and Yunlei Huang\inst{1} \and C-Y. Ng\inst{2,3} \and Lili Yang\inst{1,4} \and Sujie Lin\inst{1}
          \and
          Yihan Liu\inst{1}\fnmsep\thanks{Corresponding author: \texttt{liuyh363@mail.sysu.edu.cn} } 
          }

   \institute{School of Physics and Astronomy, Sun Yat-Sen University, No. 2 Daxue Road, 519082, Zhuhai China
         \and
            Department of Physics, The University of Hong Kong, Pokfulam Road, Hong Kong
            \and
            Hong Kong Institute for Astronomy and Astrophysics, The University of Hong Kong, Pokfulam Road, Hong Kong
            \and 
            Centre for Astro-Particle Physics, University of Johannesburg, P.O. Box 524, Auckland Park 2006, South Africa
             }

   \date{Received XXX, YYY}

 
  \abstract
   {Pulsar Wind Nebulae (PWNe) are key astrophysical laboratories for high energy phenomena. Specifically, radio observations and related polarimetry are essential probes to understand acceleration and transport, as well as PWN interaction with environment.}
   {We aim to better study the multi-wavelength morphology and magnetic geometry of \gname\ PWN (a system between early and middle ages).}
   {We conduct high resolution VLA observations at 3 cm (X band), 6 cm (C band), and 13 cm (S band) and compare them with the archival Chandra X-ray data. We also performed spectral analysis and radio polarimetry based on our radio observations.}
   {Our new VLA observations reveal a north-south double-lobed PWN bracketing a bridge-like feature, with the pulsar clearly resolved at C and S bands. The polarization fraction reaches 30\% across all bands, with the bridge region showing ordered north-south magnetic fields aligned with the X-ray torus elongation, while the southern outer lobe exhibits fields not following such a direction and the northern lobe displays a more chaotic configuration. Notably, we detect a significant radio-X-ray anti-correlation near the pulsar, with bright radio emission appearing just beyond the compact X-ray PWN boundary, multiwavelength spectral analysis suggest distinct particle populations. The radio PWN spectral index steepens from $\alpha\sim-0.3$ in the inner bridge to $<-1.0$ in the outer lobes, yet we suggest it is less likely related to synchrotron cooling. We tried to use a thick torus model with toroidal $B$-field to reproduce observed features; the result implies possible particle deceleration in the radio PWN. The equipartition magnetic field strength is estimated to be $\sim$15.3\,$\mu$G.}
   {}
    \maketitle
    \nolinenumbers
%

\section{Introduction}

A supernova explosion leaves behind a relic called supernova remnant (SNR) in the vicinity; a rapidly rotating neutron star lies at the center, observed as a pulsar. Pulsar steadily transfers its rotational energy via relativistic winds. 
Accelerated particles in the vicinity can generate synchrotron and inverse Compton emissions visible from radio to gamma-ray bands, observed as pulsar wind nebulae (PWNe). 
Though tens of PWNe have been detected, many PWNe show diverse multi-wavelength morphologies (possibly due to PWN evolution) and are yet to be well understood; besides, there are still open questions about how particles are accelerated and nebular $B$-fields form, which may help answer the common association between PWNe and TeV emissions (e.g., LHAASO) but with spatial offsets \citep{2024ApJS..271...25C}.

High resolution Chandra observations of X-ray PWNe show two types of morphologies: one-sided and equatorial torus/tori with polar jets \citep[][]{2008AIPC..983..171K,2004ApJ...601..479N}. 
The former commonly show a compact head near the pulsar and a well extended "tail" in multi-wavelength observations \citep[see][]{2017hsn..book.2159S}, which are considered as older PWN systems with a "crushed" head; the latter are believed to connect with the most accelerated particles near the termination shock (TS) in young or middle-aged systems; while the X-ray emitting particles are strongly affected by rapid synchrotron (SYN) energy loss, so more X-ray emissions are close to the pulsar.
Radio PWNe (with longer cooling time scale) can trace aged particles spread to outer space and hence a longer history of PWN-SNR co-evolution.
Middle-aged radio systems ($\tau\gtrsim$10\,kyr) commonly show double-lobed or tongue-like morphologies enveloping the X-ray PWN \citep[e.g., Boomerang, Vela
PWNe;][]{2026arXiv260220230L,2003MNRAS.343..116D} with toroidal $B$-fields, all implying a thick torus geometry \citep{2011ApJ...740L..26C}.
Young radio PWNe show different geometries; some (e.g.,  3C 58 and G54.1+0.3) are elongated with complicated detailed features (e.g., filaments, wisps) inside and rather linear $B$-fields \citep{2010ApJ...709.1125L,2002nsps.conf....1R}; while others show near spherical radio geometries with radial $B$-fields \citep[e.g., G21.5-0.9 and G292.0+1.8;][]{2022ApJ...930....1L,2003ApJ...594..326G}.
It is intriguing that radio and infrared observations of G21.5-0.9 PWN show toroidal $B$-fields close to the pulsar, transferring to a radial field further out. Yet where and how such a change happens remains unclear \citep{2022ApJ...930....1L,2012A&A...542A..12Z}.

\begin{figure*}[ht!]
    \centering
        \includegraphics[width=0.95\linewidth]{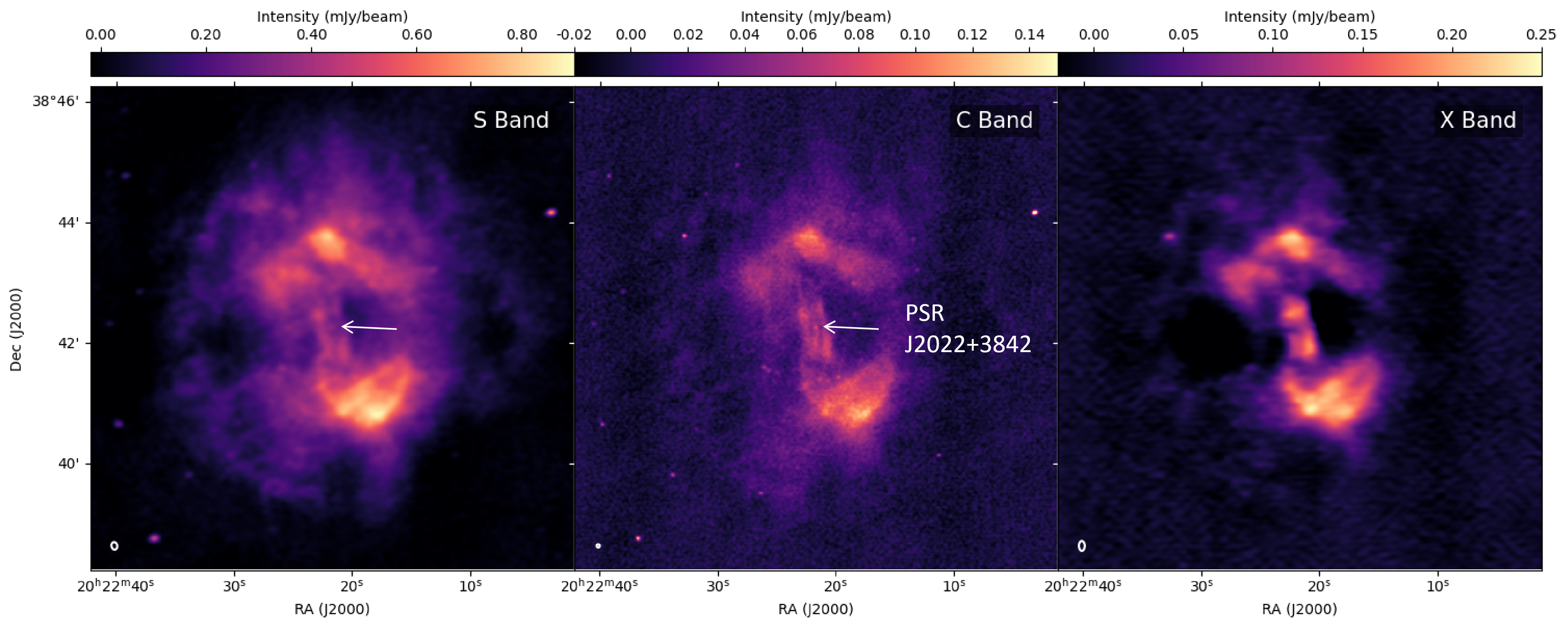}
    
    \caption{X, C, and S band total intensity maps of \gname\ PWN, with arrows marking the pulsar position and elliptical beam sizes overlaying at bottom left of each panel; the color bars show total intensity of each band with units of \mJyPerBeam.}
    \label{fig:total_I_map}
    \vspace{-1.em}
\end{figure*}

A young system \gname\ PWN interestingly shows a double-lobed radio feature commonly seen among middle-aged PWNe.
The PWN is associated with a Galactic pulsar J2022$+$3842 with a period $P\approx$48.6\,ms and period derivative $\dot{P}$ of around $8.6\times10^{-14}\,\mbox{s}\,\mbox{s}^{-1}$; 
the spin-down power $\dot{E}=3.0\times10^{37}\,\mbox{erg}\,\mbox{s}^{-1}$ and a characteristic age $\tau=8.9$\,kyr, suggest that this PWN is a rather special case between early and "middle-aged" segment and powered by a very energetic pulsar \citep{2011ApJ...739...39A}.
Chandra discovered a very compact X-ray PWN, which is elliptical and $16^{\prime\prime}\times10^{\prime\prime}$ in size, associated with PSR~J2022$+$3842 \citep{2011ApJ...739...39A},  
The radio pulsar was detected by the Green Bank Telescope (GBT) at 2\,GHz with a period of $\sim24\,\mbox{ms}$ and a large Dispersion Measure (DM) of $430\,\mbox{pc}\,\mbox{cm}^{-3}$\citep{2011ApJ...739...39A}. 
The G76.9$+$1.0 PWN was firstly observed with the Effelsberg 100-m radio Telescope; a following VLA observation resolved the double-lobed radio nebula (much larger than X-ray PWN) connected by a bridge, the bridge was firstly regarded as a background source but later found showing alignment with the X-ray PWN elongation \citep{1993A&A...276..522L,2011ApJ...739...39A,Wendker1984g76.9}. 
It is not understood why comparison interestingly shows the radio PWN much larger than X-ray structures in size \citep{2011ApJ...739...39A}, even powered by such an energetic pulsar. 
Besides, proper polarimetry measurement to understand the spatial distribution of the $B$-field is still absent.

Therefore, we here report new VLA observation results of \gname\ with higher resolution.
As illustrated, these new observations may shed new light on the puzzling morphological, spectrum and magnetic properties of not only \gname\ but also analogous systems like G21.5-0.9.    
Section \ref{sec:obs_data} includes information about observation and basic data reduction. We then report multi-wavelength observational results of this study in Section \ref{sec:results}. Some interesting results are then further discussed in Section \ref{sec:disc}, and all these studies are concluded in Section \ref{sec:concl}.
\vspace{-1em}

\begin{table}[h!]
\small
    \centering
    \begin{tabular}{llll}
    \\
    \hline
   Observation & Array   &  Frequency       &  Integration    \\
    Date        & Config. &  Coverage (MHz)  &   Time (min)     \\
         \hline
       2023 Oct. 22  & D & 7977-12025 & 85.8 \\
        2025 Jul. 21 & C & 3977-8025 & 62.3 \\
              2025 Aug. 14 & C & 1989-4013& 62.6 \\
         \hline 
    \hline
 Observation & Inst.     &    Chandra  &  Act. Exp.    \\
 Date        &           &    ObsId    &   Time (ks)    \\
    \hline
    2020 Feb. 6 \  \ & ACIS-S  & 23048 \ \ \ \ \ \ \ \  \ \ \ \ \ \ \ \ \ \  &  49.41 \ \ \ \ \ \ \ \ \ \\
    2020 Feb. 25 & ACIS-S & 23049 &  49.41 \\
    2020 Feb. 27 & ACIS-S & 21313 &  29.67 \\
    2020 Feb. 29 & ACIS-S & 23173 &  23.00 \\
    2020 Mar. 1 & ACIS-S & 21314 &  24.74 \\
    2020 Mar. 1 & ACIS-S & 23174 &  24.56 \\
    \hline 
    \end{tabular}
    \caption{VLA and Chandra observations in this study}
    \vspace{-2.5em}
    \label{tab:obs}
\end{table}

\section{Observations and Data Reduction}
\label{sec:obs_data}

\subsection{Radio Observations}

We performed new multi-band VLA radio observations of the \gname\ PWN region on  2023 October 22 at 3\,cm (X) band, then at 6 and 13\,cm (C and S) band on 2025 July 21 and 2025 August 14, respectively.
The X band observation has a frequency coverage centered at 10\,GHz, and the C and S band observations have frequency centers at 6 and 3\,GHz, respectively; while only S band observation has a frequency bandwidth of 2\,GHz, the bandwidths for the other two are 4\,GHz. 
The X band observation used the \textbf{D} configuration, while the rest observations used \textbf{C} configuration, which helps to achieve enough resolution for detailed features. 
With such observational configurations, the X band observation has a \uv\ coverage of 0.7-38.6\,k$\lambda$, while the other two cover 0.5-85.2\,k$\lambda$ in C band and 0.2-42.6\,k$\lambda$ in S band in the \uv\ space, respectively.
We used 3C286, 3C48, and 3C138 as the primary calibrators for X, C, and S band observations, respectively; and all observations used J2015+3710 as the secondary calibrator.
For the polarization calibrator in X band, we took J1407+2827, while switched to J2355+4950 in the following S and C band observations. 
The total integration times on the target are 85.8, 62.3, and 62.6\,min in X, C, and S bands, respectively.

The radio data was reduced and analyzed using the Common Astronomy Software Applications (CASA) package \citep{2022PASP..134k4501C}.
We followed a standard data reduction process to calibrate bandpass, complex gain, and flux scale of data with observations on calibrators mentioned above. We also determined the parallel-hand delay before bandpass calibration, and the cross-hand delay using single band delay method. Then the polarization (PL) calibrations were performed to extract PL properties of our target.
Most severe radio frequency interferences (RFIs) have been concerned and flagged.

All observations on the target have phase centers at the PSR\,\psr, and are sufficient to cover the whole previously detected \gname\ region \citep{1993A&A...276..522L}. 
After data examination and applying calibration to the target data, we produced full polarization dirty maps of \gname\ PWN at each band using task \texttt{tclean}, during which we used the multi-frequency synthesis algorithm as well as Briggs weighting algorithm with \texttt{robust}=0.5 to balance the sensitivity and the resolution of the image. In this task, we also performed decovolution process with interactively restricted cleaning area.

\subsection{X-ray Observations}

To understand the multiwavelength properties of this PWN, we also used archival Chandra X-ray observations for comparison. 
We only included the 6 observations in 2020 February and March. 
All observations were taken with the Advanced CCD Imaging Spectrometer (ACIS) S-array and in the full frame TIMED EXPOSURE/VFAINT mode.  
We performed data reduction with CIAO 4.16 (including sherpa) and CALDB 4.11.5. 
After obtaining data with \texttt{download\_chandra\_obsid}, we reprocessed the data with task \texttt{chandra\_repro}, generally following the CIAO science threads to analyze these X-ray data.
Further detail about data and analyses are in Section \ref{sec:results} and Table \ref{tab:obs}.

\section{Results}
\label{sec:results}

\subsection{PWN Morphologies}

The X, C, and S band total intensity results of \gname\ are shown as Figure \ref{fig:total_I_map}.
These maps have beam sizes of 10.0\arcsec$\times$6.0\arcsec, 3.4\arcsec$\times$3.0\arcsec, and 7.7$\arcsec\times$5.6\arcsec, as well as rms noise levels of 4.7, 4.2, and 10.9\,$\mu$Jy\,beam$^{-1}$, at X, C, and S bands, respectively. 
Remarkably, our high resolution radio continuum map revealed a point-like source, which can be identified as the radio and X-ray pulsar \citep{2011ApJ...739...39A} at C and S bands, while the X band observation failed to detect this source, possibly due to intrinsically much lower pulsar brightness.
In the vicinity of the pulsar (the bridge-like region), our new observations for the first time resolved more features. Two north-south linear features are discovered on the edge of the bridge in C and S bands, along the elongation of the inner feature, while it is detected to be much fainter in regions between. 
The C band map also interestingly resolves a wisp-like feature extended from the pulsar region to the south. 
The X band map has not resolved these linear features but a compact double-lobed feature (with each $\sim0.5\arcmin$ in size) also detected in some other middle-aged PWNe \citep{2003MNRAS.343..116D,2023ApJ...945...82L}, which is possibly due to larger beam size with Configuration \textbf{D}.

\begin{figure}[hb!]
    \centering
    \includegraphics[width=0.8\linewidth]{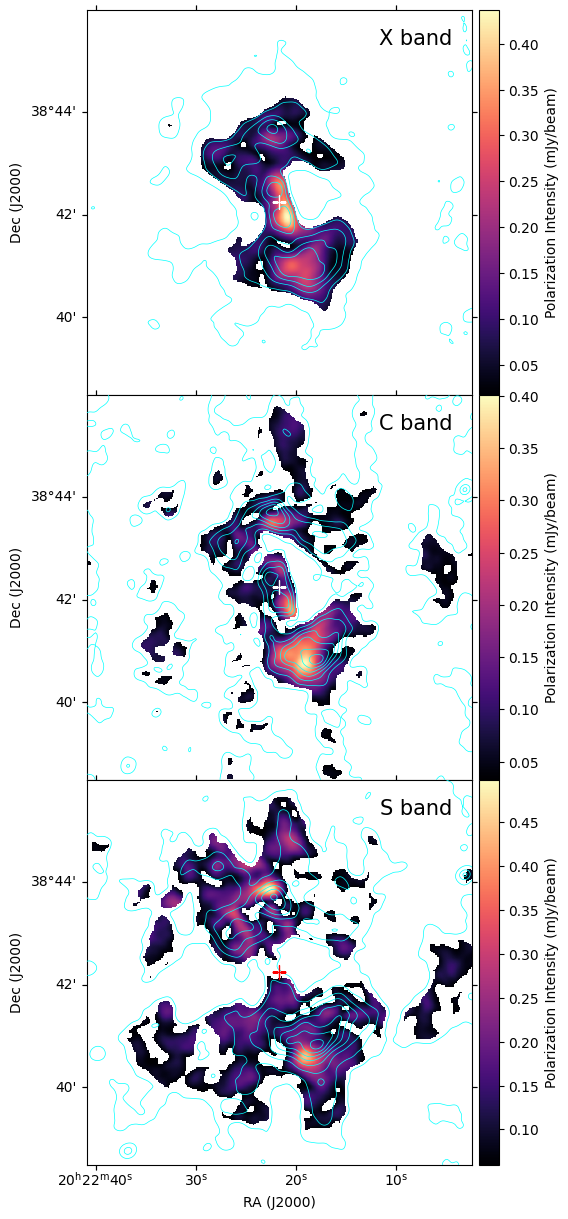}
    
    \caption{Polarization intensity (PI) maps of X (\textit{Top}), C (\textit{Middle}), and S (\textit{Bottom}) band \gname\ PWN, with crosses marking the pulsar position and color bars on the right.
    The cyan contours of total intensity maps have ranges of  0.03 to 0.8, 0.02 to 0.8, and -0.06 to 0.26\mJyPerBeam, as well as steps of 0.11, 0.11, and 0.26\mJyPerBeam, at X, C, and S bands, respectively.
    }
    \label{fig:pi_map}
\end{figure}

The large scale structures in all bands have a north-south double-lobed feature (with a diameter of $\sim$4\arcmin), bracketing a bridge-like feature with a scale of ($\sim70\arcsec\times30$\arcsec); therefore the emissions are rather faint on the eastern and western parts of the PWN. 
Both lobes have kidney-like morphologies, east-west elongated with scales of $\sim2\arcmin\times1\arcmin$; while the southern lobe is a bit more compact and brighter.
The outer lobes show more brightness than the inner bridge features, especially at lower frequency bands; taking the S band map as an example, the peaks are around 1.5\arcmin\ north and south away from the central pulsar in the lobes, reaching 1.0\mJyPerBeam, while the bridge-like region only reaches 0.6\mJyPerBeam.
Our new observations also resolve detailed structures, such as wisp-like features connecting the inner bridge with the north and south lobes in our C band map.

In our new observation, particularly in S band, a distinct circular outer boundary of the entire system is clearly detected.
This boundary, which can be identified as the shell of SNR, has a radius of $\sim3\arcmin$ and is more prominent along the eastern semicircle.
Besides, extended structures (e.g., filaments) between the PWN and the outer boundary have also been detected in S and C bands.
Neither of these, however, is significant in our X band map.
It is more likely an intrinsic property than the missing flux problem, as the \textbf{D} Configuration X band observation has sufficient \uv\ coverages corresponding to large-scale features. 

\begin{figure*}
    \centering
    \includegraphics[width=0.44\linewidth]{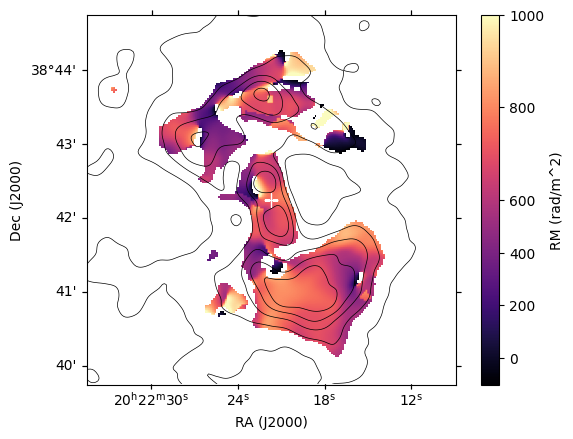}
    \includegraphics[width=0.44\linewidth]{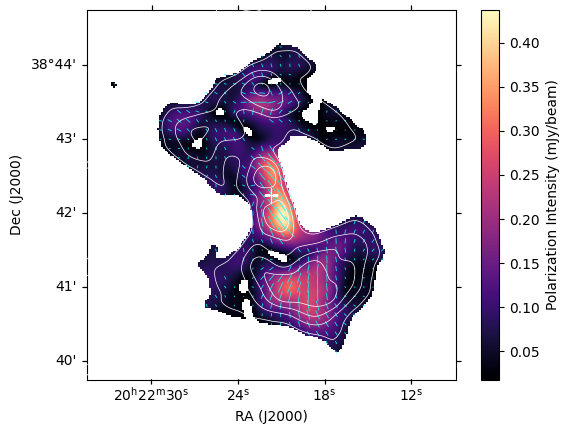}
    \caption{\textit{Left:} Rotation measurement of the \gname\ PWN, with X band intensity contour overlaid. The levels of the contours are same as that of Fig. \ref{fig:pi_map}. The pulsar position is marked by a white cross. \textit{Right:} Faraday corrected $B$-vector map (90$^\circ$ rotation from original PA map) overlaying the X band PI map and Stokes \textit{I} contours as in Figure \ref{fig:pi_map}.}
    \label{fig:RM_Bvec}
\end{figure*}

\subsection{Polarization Intensity Maps}

Figure \ref{fig:pi_map} shows polarization intensity (PI) maps of \gname\ PWN at X, C, and S bands. 
Similar to what is illustrated in Section \ref{sec:obs_data}, we produce full polarization cleaned images in all bands, but smoothed to a uniform beam size with a full width half maximum (FWHM) of 30\arcsec\ to optimize the Signal to Noise ratio (S/N).  
We then use the CASA task \texttt{immath} to plot the PI map in each band, based on Stokes \textit{Q} and \textit{U} clean maps.   
The rms noise of Stokes \textit{Q/U} maps are 11.3, 26.3, and 6.0\,$\mu$Jy\,beam$^{-1}$ at S, C, and X bands, respectively; we then clip the pixels where both PI and total intensity have S/N\textgreater3 in all bands. The Ricean bias is not removed to maximize the polarization intensity.

The \gname\ PWN is highly polarized in each band, with linear polarization fraction (PF) of 38.4$\%$ in X band, 44.1$\%$ in C band, and 31.5$\%$ in S band. The polarized emission of the PWN generally follows the total intensity, but shows difference in some detailed regions. 
The C and X band polarized emissions show similar geometry, with the bridge in the center corresponding to the same structure in the total intensity map, connecting the two northern and southern outer lobes.
In these two bands, the polarization emissions of the bridge generally show a double-lobe structure like the Stokes \textit{I} geometry in the bridge, while slightly more PL emissions are on the west side. 
The two outer lobes of the PWN also show obvious bright polarization emission, but the emission regions are not as extended as those in the total intensity map.
The southern outer lobe is brighter and more diffuse, while the northern lobe is dimmer and less polarized. 
The overall polarized emission of the S band PWN is more extended to the SNR Rims than the other two bands. However, it only peaks at two compact regions at $\sim$1\arcmin\ north and south from the center pulsar, near the extended line of the bridge, and does not show a bright emission at the central bridge region. Since our S band observation has a wide bandwidth coverage of 2\,GHz at long wavelength, this could be the result of bandwidth depolarization as detected in G11.2-0.3 PWN \citep{2025ApJ...988..163Z}.

\begin{figure*}[]
    \centering
    \includegraphics[width=0.8\linewidth]{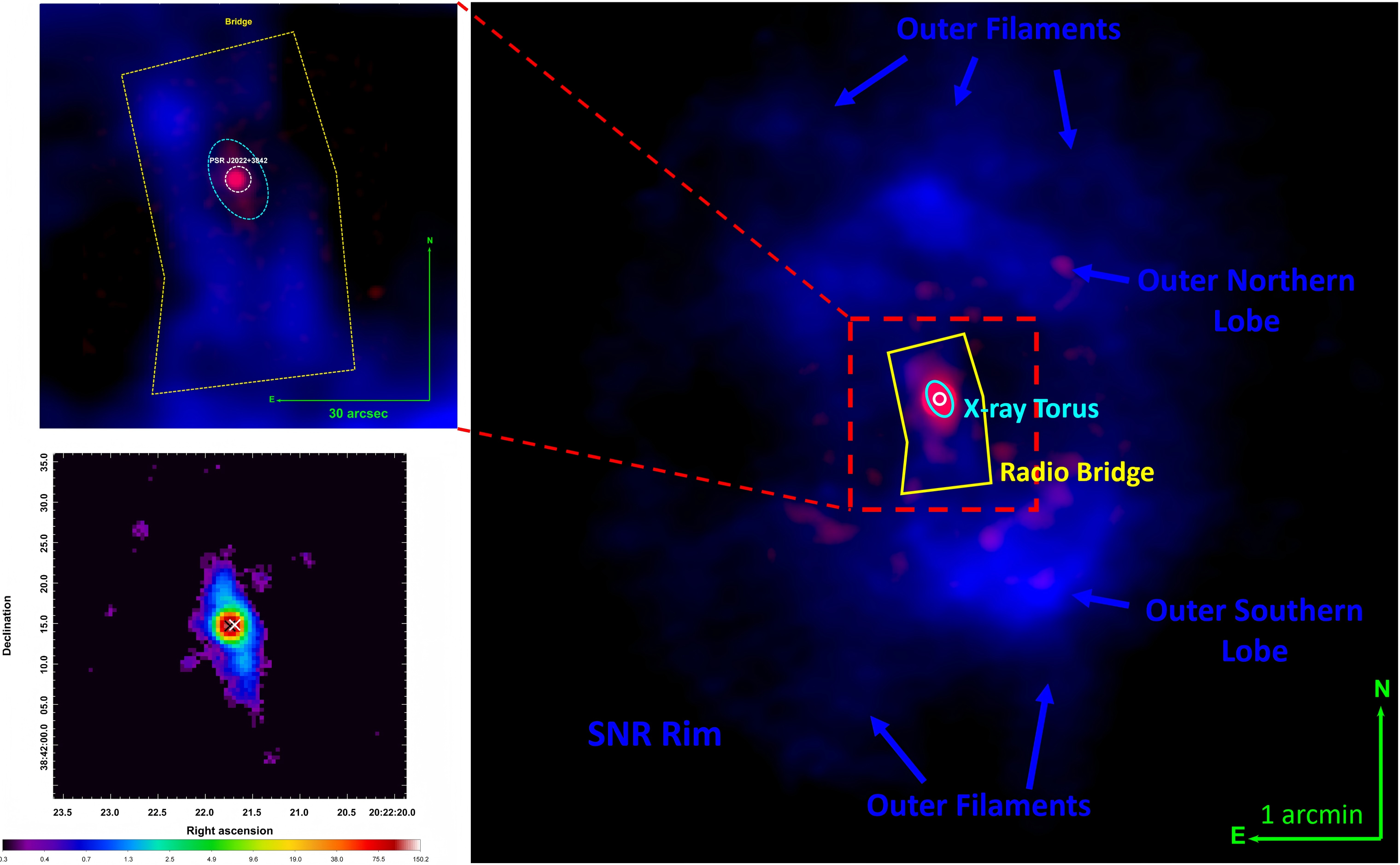}
    \caption{\textit{Bottom Left}: 3-9\,keV Chandra X-ray map of \gname\ X-ray torus, the white and black crosses show previous and current pulsar positions. \textit{RGB Images}: Comparison of S band radio (blue) and X-ray (red) maps of \gname\ PWN, with the yellow polygon, cyan ellipse, and white circle in both images indicating to regions of the radio "bridge", X-ray torus, and the pulsar, as regions for spectral analyses.   }
    \label{fig:radio_X-ray}
\end{figure*}

\subsection{Rotation Measure and Magnetic Field Configuration}

Though SYN emission original polarization angle (PA) is perpendicular to local $B$-field, Faraday effect in foreground medium leads to PA changes $\Delta\chi$ as,
\begin{equation}
    \Delta\chi = RM\cdot\lambda^2,
\end{equation}
where $RM$ is the rotation measure of Faraday effect, and $\lambda$ is the wavelength.
We then try to obtain the local RM distribution of the PWN by linearly fitting the PA maps in different wavelengths. 
To obtain this, we first plot PA maps of each band with the task \texttt{immath}.
One significant problem in the linear fitting for RM is the $n\pi$ ambiguity, as the polarization direction is periodic. 
We do not employ the CASA task \texttt{rmfit}, which shows limited performance in our case, but develop a new logic, firstly searching for regions with both significant S/N and good fitting as references, and then attempting to "correct" the PA values according to the nearest reference point. 
The RM map after fitting is shown as Figure \ref{fig:RM_Bvec} (left panel).

The RM in the PWN generally spans from around 200 to 1000\RMu. 
RM in the bridge (showing the best S/N and fitting performance) and the southern lobe shows a uniform distribution around 700\RMu;
The fluctuation of RM is more significant in the northern lobe.
The RM drops to around 600\RMu\ in the east of the northern lobe; the western side of the northern lobe drops to around 0\RMu, which is possibly related to not fully eliminated $n\pi$ ambiguity issues.

We also obtain the Faraday corrected PA map from our RM fitting algorithm as illustrated above, then plot the $B$-vector map, as shown in Figure \ref{fig:RM_Bvec}.
The $B$-vectors in the bridge region are generally in the north-south (NS) direction along its elongation. 
In the outer southern lobe, $B$-vectors are not generally ordered but show a potential trend perpendicular to the vectors in the bridge; fields in the northern lobe become rather chaotic, for example, east-west $B$-vectors (parallel to the northern lobe) are detected close to the PI peak in the northern lobe, while $B$-field in other regions are still following that in the bridge region.  

\begin{figure}[hb!]
    \centering
    \includegraphics[width=0.8\linewidth]{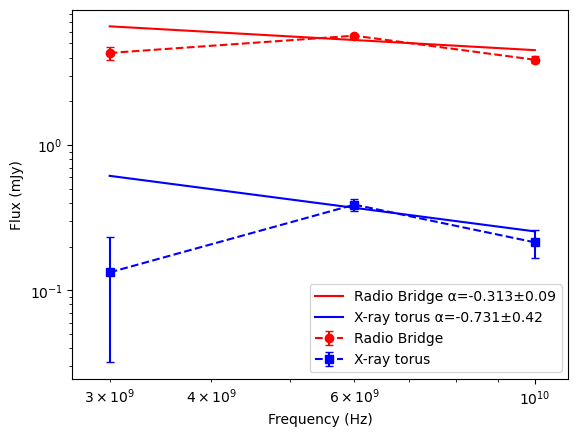}
    \includegraphics[width=0.8\linewidth]{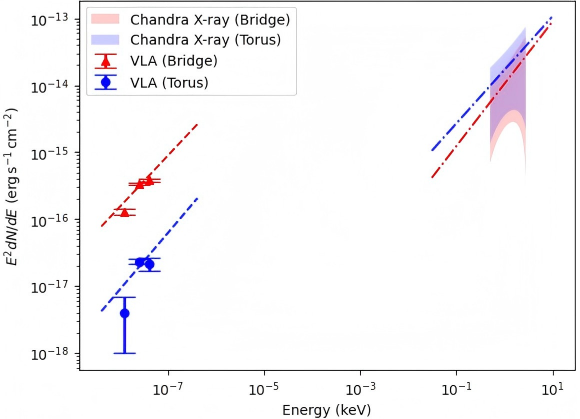}
    \caption{\textit{Top}: Radio spectra of X-ray bright (torus) region and radio bridge region in the \gname\ PWN. \textit{Bottom}: Multi-wavelength spectra of these two regions, with extrapolated spectra shown in dashed lines. } 
    \label{fig:radio_spec}
\end{figure}

\subsection{Radio and X-ray Comparison}
\label{subsec:radioX-rayComparison}

Figure \ref{fig:radio_X-ray} compares the Chandra X-ray images (at 3-9\,keV) and S band radio map of the whole \gname\ region, and the region close to the pulsar. 
The bottom left panel shows that the compact X-ray PWN generally aligns with the previous observation \citep{2011ApJ...739...39A}; the torus is NS elongated and there are some fainter structures detected extending to the east and west, which are considered to be the jets \citep{2011MNRAS.416.2560M}.  
For the large scale comparison over the whole SNR, no significant X-ray emissions are detected coinciding with the outer lobes and SNR, but much fainter X-ray emissions ($\sim1.5\arcmin$) extending into the radio "bridge" region.
In the center, a point-like X-ray source can be identified as the pulsar, and its counterpart in radio is also discovered in our radio observation.
For the PWN region, significant radio and X-ray brightness anti-correlation is interestingly detected in the bridge region.
The compact X-ray PWN is north-south elongated to around 10\arcsec\ away,  exactly beyond which radio emission becomes bright.

\subsection{Radio and X-ray Spectra}

To investigate the spectral properties close to the pulsar, we analyze the radio and X-ray spectra in X-ray "torus" region and radio "bridge" region.
As no significant morphological changes are found between the new and previous Chandra X-ray images, we here regard that the X-ray emissions consists of a point-like pulsar and a compact PWN.

Accordingly,  we use a circle with a 2.5\arcsec\ radius as the region of pulsar, and a $8.3\arcsec\times5.1\arcsec$ ellipse (with a position angle of -64$^\circ$) beyond the pulsar as the X-ray torus region \citep{2011ApJ...739...39A}; besides, we draw a bridge-like polygon as shown in Figure \ref{fig:radio_X-ray}, and the region between it and the X-ray ellipse is considered as the region of the radio "bridge". 
For the background region, we also set an annulus with radius of 20\arcsec\ and 40\arcsec\ centered at the position of pulsar where is free of bright or extended emissions.

Then we extract the non-thermal X-ray spectra of both the radio and X-ray bright PWN regions with the CIAO task \texttt{specextract}; it is notable that the radio and X-ray PWN regions in our estimate are dually exclusive, as is illustrated above.
Next, we use Sherpa to fit the properties of these X-ray spectra between 1.5 and 7.5\,keV.
Every 20 counts are grouped and the background has been subtracted in our analyses. 
Based on previous observational results, we choose a model of \texttt{xsphabs$\times$powlaw1d} and for the fit of the inner X-ray torus we get a column density of  N$_H$=2.24$\pm$1.06$\times10^{22}$\,cm$^{-2}$, and then fix it in the following fitting of the radio bridge region.
Our final results show photon indices $\Gamma$ of 1.2$\pm$0.5 and 1.1$\pm$0.6 in the X-ray torus and radio bridge regions, as in Figure \ref{fig:radio_X-ray}, respectively; and it is notable that the result is aligned with the previous result \citep{2014ApJ...790..103A}.
The unabsorbed X-ray fluxes (X-ray torus and radio bridge) at 0.5-7\,keV are 3.2 and 2.6$\times$10$^{-14}$\,erg\,cm$^2$\,s$^{-1}$.
All these are listed in Table \ref{tab:sed}.

We also calculate the radio spectra of the same regions in the "bridge". 
We use the same circular region of the X-ray pulsar referring from the previous Chandra X-ray analysis. 
As SNR features are overlaying with the PWN, we select several background regions in the SNR region, rather than totally source-free regions, to mitigate the SNR component contamination in our spectral analysis. 
We check and find that no significant radio pulsar emission in C and S band is detected beyond the X-ray circle of pulsar; the X band observation did not resolve the pulsar from the PWN, possibly due to lower brightness of the pulsar at X band and the slightly larger beam size as mentioned. 
Our measurement gives radio pulsar flux densities of 0.09$\pm$0.01 and 0.04$\pm$0.04\mJyPerBeam\ at C and S bands, respectively.

Figure \ref{fig:radio_spec} and Table \ref{tab:sed} show the integrated flux of the regions as above. 
X-ray bright regions (torus) close to the pulsar have flux densities of 0.21$\pm$0.05, 0.39$\pm$0.04, and 0.13$\pm$0.10\,mJy at X, C, and S bands, respectively; for the radio bright region (bridge), flux densities are 3.86$\pm$0.21, 5.65$\pm$0.16, and 4.30$\pm$0.45\,mJy at X, C, and S bands, respectively. 
Though we fit the radio spectra of these regions with a single power law, both regions cannot be well fitted with such a distribution; in other words, the radio spectra in the bridge peak at C band. 
In any case, the single-power-law fitting shows radio spectral indices of -0.73$\pm$0.42 and -0.31$\pm$0.09 in the X-ray bright and radio bright regions. 

\begin{table}[h!]
   
    \centering
    \small
    \begin{tabular}{lll}
    \hline
       Radio bands\tablefootmark{a} & X-ray "Torus"  & Radio "Bridge" \\
       \hline
        S band &  0.133$\pm$0.10  & 4.30$\pm$0.45 \\
        C band &  0.389$\pm$0.04 &  5.65$\pm$0.16 \\
        X band &  0.214$\pm$0.05 & 3.86$\pm$0.21\\\hline
    X-ray Spectra \\
    \hline 
    $\Gamma$ & 1.20$\pm$0.50 & 1.07$\pm$0.59 \\
    Amplitude\tablefootmark{b} & 6.61$\pm$5.24 & 4.09$\pm$3.19 \\
    $N_H$\tablefootmark{c}  & 2.25 & 2.25 (frozen) \\
    $\chi^2$/DoF & 10.6/23 & 22.7/66 \\
    \hline
    \end{tabular}
    \caption{Spectral properties of \gname\ PWN bridge regions. }
    \raggedright
    \tablefoot{The radio fluxes are in the upper half panel and X-ray spectra are in the lower half.}
    \tablefoottext{a}{in a unit of mJy.\\}
    \tablefoottext{b}{in a unit of 10$^{-6}$.\\}
    \tablefoottext{c}{in a unit of 10$^{-22}$\,cm$^{-2}$.}
    \vspace{-0.6em}
     \label{tab:sed}
    
\end{table}

\subsection{Spectral index map}

The variation in morphology of \gname\ PWN between different frequencies indicates its spectral variation throughout the PWN. To better understand the nature of its spectrum in different regions, we calculate the radio spectral index map of \gname\ PWN between the S, C and X bands pixel by pixel. We use a single power-law spectrum to describe the PWN: \(S \propto \nu^{\alpha}\), where \(S\) is the flux density, \(\nu\) is the frequency and \(\alpha\) is the spectral index. 
During our calculation, we filter the three bands of data to have the same \uv\ coverage of 0.8-30.0\,k$\lambda$ so that they have the same missing flux problem. To obtain an image with better S/N, we smooth the total intensity images of the three bands using the same \uv\ taper of 10\arcsec\ FWHM. 

\begin{figure}[ht!]
    \centering
    \includegraphics[width=0.85\linewidth]{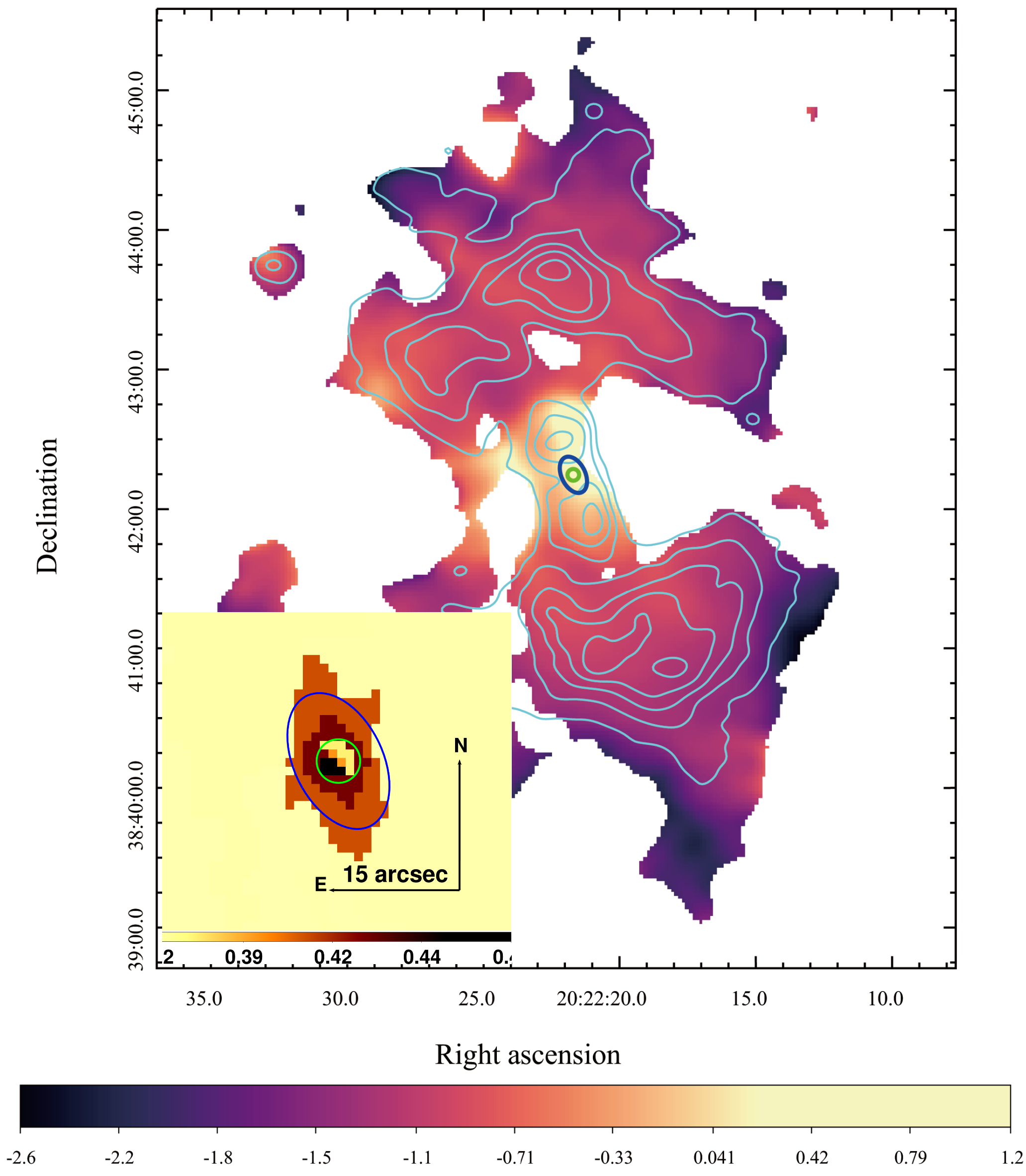}
    \caption{Radio spectral index map of the \gname\ PWN with overlaid contours of X band total intensities of 0.025, 0.070, 0.115, 0.205, 0.250\mJyPerBeam. The inset image shows spectral hardness of the X-ray torus. Both images are overlaid with the ellipse and circle representing X-ray torus and pulsar same as Figure \ref{fig:radio_X-ray}.\vspace{-1em}}
    \label{fig:radio_spix_map}
\end{figure}

The final spectral index map, shown in Figure \ref{fig:radio_spix_map}, is masked if the error of the spectral index fitting is larger than 0.7. 
The spectral index of \gname\ PWN has a roughly north-south symmetrical distribution, gradually decreasing from the center bridge towards the outer two lobes. Near the PWN center, the mean spectral index of the bridge is over 0; in the outer lobes the spectral index gradually drops to -1.0, while in the outer regions the spectral index eventually goes to less than -2.0. The result reveals a spatial gradient of the spectral index, which indicates to hard radio spectra in the inner part and the spectra steepening outward, 
though the maps may still experience some missing flux issues.

Additionally the inset image of Figure \ref{fig:radio_spix_map} shows the X-ray hardness map in the X-ray torus region. 
We use the \texttt{contbin} package for binning of the X-ray image with intensity \citep{2006MNRAS.371..829S}.   
Unfortunately, the regions are too compact for X-ray spectral fitting, therefore we produce 0.5-2.5 and 2.5-7.5\,keV images and calculate the hardness map with the task \texttt{dmimgcalc}, dividing their difference map by their sum map. 
Apart from the pulsar region, the hardness drops when moving away from the pulsar, which generally aligns with the trend of spectral softening in the outer region.

\section{Discussion}
\label{sec:disc}

\begin{figure*}[ht!]
    \centering
    \includegraphics[width=1\linewidth]{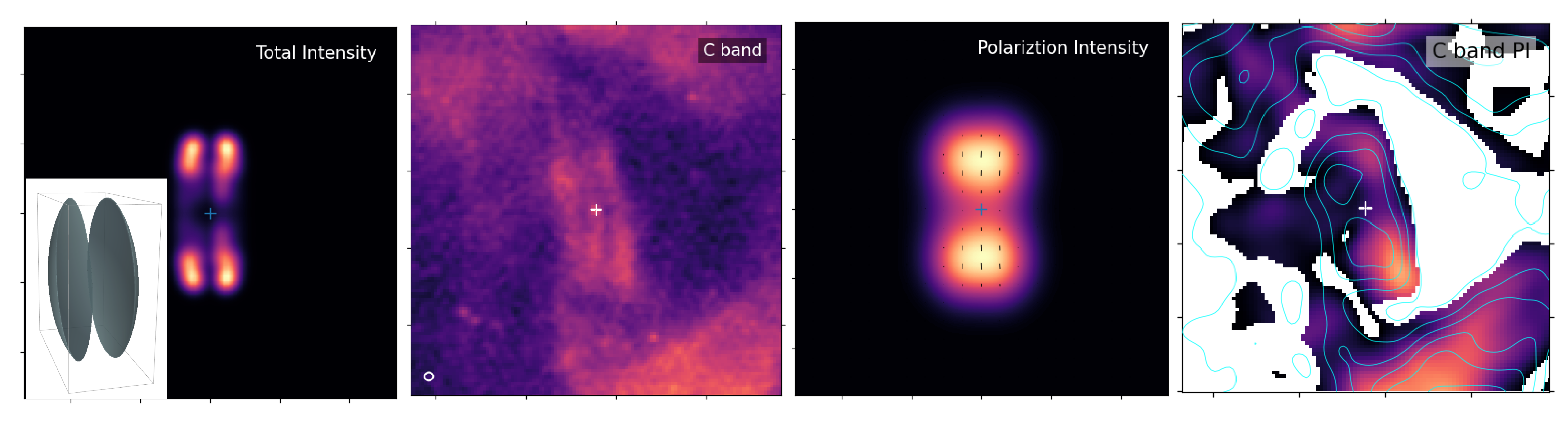}
    \caption{Model of \gname\ PWN, including the simulation of total intensity, polarization intensity, and magnetic field geometry, inset image of the total intensity simulation showing the geometry of the model. The corresponding C band observation result is on the right to compare.}
    \label{fig:plmodel}
\end{figure*}

\subsection{Multi-wavelength Morphology in \gname}

The radio and X-ray anti-correlated feature (as shown in Section \ref{subsec:radioX-rayComparison}) has been detected in several other middle-aged PWNe (e.g., Vela and B1706 PWN), 
that compact X-ray torus/jet PWN features are only detected close to  the pulsar and the radio PWN becomes bright beyond X-ray emissions \citep{2003MNRAS.343..116D,2023ApJ...945...82L}.   
One possible explanation is rather low number density of relativistic particles in X-ray emitting regions, while cooled-down particles (which no longer generate X-rays) pile up when spreading to the outer region and become bright in radio \citep{2008ApJ...684..542K}. 
Meanwhile, some theories like magnetic reconnection within the pulsar wind suggest the intrinsic spectral break of accelerated particles in PWNe \citep{2011ApJ...741...39S}; hence the radio emission close to the acceleration sites could be much fainter than the X-ray spectral extrapolation. These spectral features have been recently detected in the multiwavelength observations of several PWNe like G11.2-0.3 \citep[][]{2025ApJ...988..163Z}.

To further confirm this, we combined the radio and X-ray spectra to constrain the multiwavelength synchrotron spectral energy distribution (SED) of the inner bridge regions. All these results are shown in the lower panel of Figure \ref{fig:radio_spec}. 
It is very interesting to see that the radio and X-ray spectra cannot intersect well for either the X-ray torus or the radio bridge region. 
We note that the Chandra X-ray results have hard spectral indices $\Gamma$ close to 1.0, which coincides with previous detection but suggests a so faint counterpart in radio \citep{2014ApJ...790..103A}. 
If so, the multi-wavelength spectra suggests 
that the outer radio structure and the compact X-ray PWN (with rather significant boundaries) are of different particle populations. 
We note that the X-ray spectra are not strongly constrained, as mentioned above. More exposure time or following UV/X-ray observations could together be helpful in discriminating the existence of intrinsically different populations. 
Nonetheless, neither scenario conflicts with the piling-up hypothesis or the spectral-break hint of SYN particles.

Figure \ref{fig:radio_spix_map} shows a common steepening trend of radio and X-ray spectra as one moves away from the pulsar; especially, the radio spectral indices show clear drops near the edge of the radio bridge (double-lobed feature). 
To test if SYN cooling dominates such steepening trend of radio spectral indices, we then estimate the equipartition $B$-field strength of the radio bridge region and then predict the SYN cooling time scale.

The equipartition magnetic field can be estimated as:
\begin{equation}
    B_{eq} = [6\pi(1+k)c_{12}L_{syn}\Phi^{-1}V^{-1}]^{2/7},
\end{equation}
where \(V\) is the source volume, \(L_{syn}\) is the synchrotron luminosity, the filling factor of emission \(\Phi\) (usually taken as 1) is the ratio between emission volume and the magnetic field volume, \(k\) is the ratio between electron energy and proton energy (usually taken as 0), and \(c_{12}\) is a constant related to frequency range and spectral index \citep{2004IJMPD..13.1549G}. To estimate the volume of radio \gname\ PWN, we consider the structure as the model we discussed in Sector \ref{subsec: PL model}, with $a=33.80$\arcsec, and $b=14.49$\arcsec.
Adopting a distance of 10$d_{10}$\,kpc \citep{2011ApJ...739...39A}, we give a volume of \(V = 2.60 \times 10^{56}d_{10}^3\,\mathrm{cm}^3\). We estimate the synchrotron luminosity of \gname\ by assuming a 
broken power law 
from \(10^7\) to \(10^{11}\)\,Hz, giving a luminosity of 
\(L_{syn} = 1.96 \times 10^{31}d_{10}^2\,\mathrm{erg\,s}^{-1}\).
These give the equipartition magnetic field strength of \(B_{eq}\approx 15.3\, d_{10}^{-2/7}\,\mu\mathrm{G}\). This is not far from a typical $B$-field strength in PWNe ($\sim$10\,$\mu$G).

Given such an equipartition $B$-field strength in the radio bridge, the time scale $\tau_{syn}$ introducing a cooling break ($\nu_b$) at GHz frequencies is:

\begin{equation}
    \tau_{syn} = (\frac{\nu_b}{10^{21}\,\mathrm{Hz}})^{-1/2}(\frac{B}{1\,\mu\mathrm{G}})^{-3/2}\,\mathrm{kyr} \gtrsim 
5284\,\mathrm{kyr},
\end{equation}
which is much larger than the characteristic pulsar age of 8.9\,kyr.
This result, if valid, rules out radiative cooling as the primary driver of the radio spectral steepening. 
It is worth noting that outer lobes are detached from the inner bridge and their $B$-field configurations do not follow that of the bridge; though not conflict with the scenario that X-ray torus powers the outer lobes \citep{2011MNRAS.416.2560M}, all these object direct particle transporting to outer regions.
In any case, given the 10\,kpc distance \citep{2011ApJ...739...39A}, the radio PWN show an average expanding velocity (equatorial) of $\sim$150\,km/s, hinting that the relativistic outflows from in X-ray PWN may experience significant deceleration and pile-up inside the "bridge" region.

Beyond the radio bridge region, the steeper radio spectra of outer lobes in the north and south could be related to older particle populations or less efficient acceleration (e.g., with SNR evolution).  
Compared with the inner bridge, the rather complex radio morphology and $B$-field configurations of the outer northern lobe, as well as the fainter polarized emissions inside, imply that particles in the outer northern lobe is running into turbulence, possibly due to interactions with the SNR (e.g., reverse shock). 
Filamentary structures between the SNR rim and PWN also suggest existence of Rayleigh-Taylor instabilities in PWN-SNR interactions.

\subsection{Local magnetic field features}
\label{subsec: PL model}

It is suggested that a model of thick radio torus can reproduce the double-lobed or tongue-like radio PWN morphologies  \citep{2011ApJ...740L..26C,2023ApJ...945...82L}, by considering relative thickness projecting to the 2D sky map and Doppler boosting.
Many of \gname\ results (e.g., radio and X-ray feature; $B$-field) suggest that its morphology could be an analogy of a torus model. We here attempt to use a toy model of NS-elongated torus to understand PL geometry of the "bridge" region.

We first consider a simple tilted torus analog feature in a 3D space.
Though it reproduces double-peaked Stokes \textit{I} morphology and NS elongated $B$-field as is illustrated,
 such a simple model also predicts PI minimum in the northern and southern ends, as varying magnetic field geometries along any single line of sight cause depolarization, conflicting with observed PI morphologies showing significant PL emissions at the ends.

We accordingly consider a toy model of hollow and oblate ellipsoid, with major radius ($a$) of $33.80$\arcsec\ and minor radius $b=3a/7$, and a subtracted inner ellipsoid with semi-axes of $(5a/2,5a/2,b/2)$. Since the semi-major axis of the subtracted ellipsoid is larger than that of the outer one, the model ends up an analogy of a double-parallel-torus similar to Vela PWN in X-ray \citep{2003MNRAS.343..116D,2004ApJ...601..479N}. Besides, the polar axis of our double-torus (aligning with the minor axis) tilted to the east from the line of sight with an angle (i.e., the viewing angle) of $84^{\circ}$. We also added Doppler boosting effect amplifying emission when particles move towards the observer and vise versa. Following \citet{1987ApJ...319..416P}, we have apparent intensities of 
\begin{equation}
    I \propto (1-n \cdot \beta)^{-(1+\Gamma)}I_0,
\end{equation}
where \(n\) is the unit vector of line of sight, \(\beta=v/c=0.2\) is the bulk velocity of the post-shock flow, \(\Gamma\) is the photon index in the rest frame, and \(I_0\) is the intensity of synchrotron emission.
For magnetic geometry, we define a toroidal $B$-field configuration about the tilted polar axis as in other PWNe.
We also introduce an emissivity ($I_0$) distribution along the  torus radial directions in our toy model, with a steep gradient $I_0 \propto R^{20}$.
Then we calculate the polarization direction in the 2D sky map and subsequently Stokes $Q$ and $U$ for each pixel in the space, which are accumulated along the line of sight direction and then used to extract the total intensity map, PI map, and observed $B$-field directions, all these are shown in Figure \ref{fig:plmodel}.

Our model reproduces the observed geometries in both total intensity and PI map. In total intensity map, the double-torus reproduces the parallel linear features in the "bridge" and the fainter gap between. 
The PI model, after smoothed by a gaussian kernel with $\sigma=3$, aligns with the observed PI map with two peaks at the NS ends and more PL emission on the west.  
Despite of simplicity, this toy model demonstrates how observed geometries require the deceleration/piling-up effect in the bridge region, consistent with a transition to slower and confined outflows as discussed above.
If so, \gname\ can not only be an inspiring case for understanding transforming PWN $B$-field geometries (e.g., G21.5-0.9), but also may represent an early stage of the thick-torus development seen in middle-aged PWNe.
Besides, this model can not fully preclude the possibility that the elongation of the "bridge" is not the torus but jets; if so, the linear $B$-field configurations along the jets are also seen in the 3C~58 and G54.1+0.3 PWNe \citep{2002nsps.conf....1R,2010ApJ...709.1125L}. 

\section{Conclusion}
\label{sec:concl}

We present study of \gname\ PWN with our new high-resolution VLA radio observations at X, C, and S bands, together with archival Chandra X-ray data. Our main findings are:

\begin{itemize}
    \item  The PWN shows a double-lobed feature bracketing a central bridge, with detailed structure and the pulsar resolved; a SNR rim is detected beyond the PWN.
    \item The radio PWN is anti-correlated to the X-ray torus, only become bright beyond it. Radio/X-ray SEDs imply distinct particle populations. The spectral index steepening trend is common in radio and X-ray PWN, while SYN cooling may not be the dominating factor.
    \item  The nebula is highly polarized: the bridge region shows ordered north–south fields aligned with the X-ray torus, while the southern lobe exhibits perpendicular fields and the northern lobe shows a chaotic configuration, suggesting a toroidal $B$-field near the pulsar becoming more turbulent outside.
    \item  A toy torus model try to reproduces the observed bridge morphology and $B$-field, implying either extreme particle pile-up and deceleration.
    An equipartition estimate predicts $B\sim$ 15.3\,$\mu$G near pulsar.
\end{itemize}
G76.9+1.0 is a rare PWN bridging young and middle-aged systems. Our results demonstrate the critical radio polarimetry and spectral properties to understand PWN magnetic evolution and particle transport. Future deep X-ray and multi-epoch observations will help discriminate between the proposed scenarios.

\begin{acknowledgements}
C.-Y. N. is supported by a GRF grant of the Hong Kong Government under HKU 17304524. 
This work is also supported by the National Natural Science Foundation of China (NSFC) grants 12261141691.
The National Radio Astronomy Observatory and Karl G. Jansky Very Large Array (VLA) are facilities of the National Science Foundation operated under cooperative agreement by Associated Universities, Inc. This research has made use of data obtained from the Chandra Data Archive and software provided by the Chandra X-ray Center (CXC) in the application packages CIAO and Sherpa.
\end{acknowledgements}

\bibliographystyle{aa_bkp}
\bibliography{g76.9}

\end{document}